\documentclass[prb,showpacs,twocolumn,amsmath,amssymb,superscriptaddress]{revtex4-1}

\usepackage{color}
\usepackage{graphicx}
\usepackage{tabularx}
\newcolumntype{L}[1]{>{\raggedright\arraybackslash}p{#1}} 
\newcolumntype{C}[1]{>{\centering\arraybackslash}p{#1}} 
\newcolumntype{R}[1]{>{\raggedleft\arraybackslash}p{#1}} 

\definecolor{kg}{rgb}{.1,0.6,.1} 

\begin{document}

\title[Features]
{Parallel carbon nanotube quantum dots and their interactions}

\author{Karin Go\ss}
\altaffiliation[Current address: ]{1. Physikalisches Institut, Universit\"at Stuttgart, Pfaffenwaldring 57, 70550 Stuttgart, Germany}
\email{karin.goss@pi1.physik.uni-stuttgart.de}
\affiliation{Peter Gr\"unberg Institut (PGI-6), Forschungszentrum J\"ulich \& JARA J\"ulich Aachen Research Alliance, 52425 J\"ulich, Germany}
\author{Martin Leijnse}
\affiliation{Center for Quantum Devices, Niels Bohr Institute,
University of Copenhagen, Universitetsparken 5, 2100 Copenhagen,
Denmark}
\author{Sebastian Smerat}
\affiliation{Physics Department, Arnold Sommerfeld Center for Theoretical Physics, Ludwig-Maximilians-Universit\"at M\"unchen, 80333 M\"unchen, Germany}
\author{Maarten R. Wegewijs}
\affiliation{Peter Gr\"unberg Institut (PGI-2), Forschungszentrum J\"ulich \& JARA J\"ulich Aachen Research Alliance, 52425 J\"ulich, Germany}
\affiliation{Institute for Theory of Statistical Physics, RWTH Aachen, 52056 Aachen, Germany}
\author{Claus M. Schneider}
\affiliation{Peter Gr\"unberg Institut (PGI-6), Forschungszentrum J\"ulich \& JARA J\"ulich Aachen Research Alliance, 52425 J\"ulich, Germany}
\author{Carola Meyer}
\affiliation{Peter Gr\"unberg Institut (PGI-6), Forschungszentrum J\"ulich \& JARA J\"ulich Aachen Research Alliance, 52425 J\"ulich, Germany}

\date{\today}

\begin{abstract}
We present quantum transport measurements of interacting parallel quantum dots formed in the strands of a carbon nanotube rope. In this molecular quantum dot system, transport is dominated by one quantum dot, while additional resonances from parallel side dots appear, which exhibit a weak gate coupling. This differential gating effect provides a tunability of the quantum dot system with only one gate electrode and provides control over the carbon nanotube strand that carries the current. By tuning the system to different states we use quantum transport as a spectroscopic tool to investigate the inter-dot coupling and show a route to distinguish between various side dots. By comparing the experimental data with master equation calculations, we identify conditions for the tunneling rates that are required in order to observe different manifestations of the inter-dot coupling in the transport spectra.
\end{abstract}
\maketitle

\section{Introduction}

Carbon nanotubes (CNTs) are a versatile material for electronics. In addition to extraordinary electronic \cite{White1998,Wildoer1998,Kong2001,Liang2001} and thermal\cite{Kim2001,Fujii2005,Balandin2011} transport properties, they are mechanically flexible and strong \cite{Dai1996,Walters1999,Yu2000}. The current is carried by the $\pi$-orbitals of the macromolecule and thus, the electronic transport can be strongly influenced by the environment. This feature can be exploited by using carbon nanotubes as the functional element in detectors, e.\,g., gas sensors\cite{Kong2000,Kawano2007}. A different way of using this property is the functionalization of individual CNTs with molecules to create new hybrid types of nano-devices such as biosensors \cite{Besteman2003,Zhang2007} or spin valves \cite{Urdampilleta2011}.

The interactions involved in such a functionalization are not yet fully understood but are expected to also play an important role in the transport properties of other interacting $\pi$-systems such as graphene or individual molecules. A lot of effort is put into studying quantized transport in these systems.\cite{Park2000,Park2002,Liang2002b,Kubatkin2003,Ponomarenko2008,Molitor2009,Champagne2005,Pasupathy2005,Heersche2006,Jo2006,Osorio2007,Trauzettel2007,Sols2007,Stampfer2008} In contrast to graphene and individual molecules, quantum transport on clean, individual CNTs today is well established and understood.\cite{Tans1997,Nygard2000,Cobden2002,Sapmaz2005,Kuemmeth2008,Steele2009b} Therefore, it can be used as a spectroscopic tool for the investigation of more complex multi-component devices.  Within this context, CNTs bundled together in form of a rope represent a generic and readily available system to study the electrical transport of interacting molecules.

Recently, we investigated the electronic hybridization between the parallel quantum dots (QDs) in a CNT rope system, when two electrochemical potentials of the dots are in resonance.\cite{Goss2011a} In this article, we focus instead on the differential gating effect and the off-resonant effects. We investigate and discuss the influence of the relative tunnel rates of the system and the coupling of the QDs to the leads.

\begin{figure}[tb]
  \includegraphics{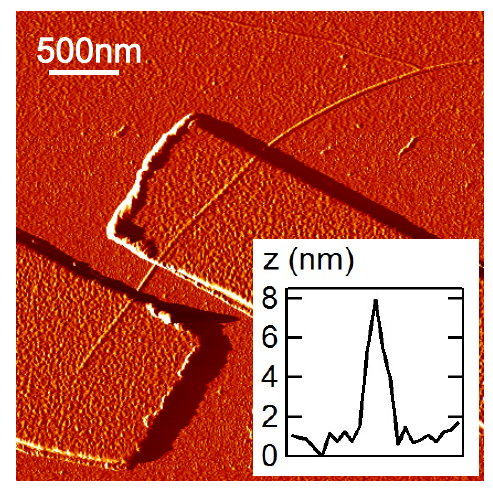}
  \caption{(Color online) Atomic force micrograph of the parallel quantum dot device with gold contacts patterned on top of a CNT rope. Inset:~Height profile $z$ in the quantum dot region between the contacts.}
  \label{fig:afm}
\end{figure}

\begin{figure*}[htb]
  \includegraphics{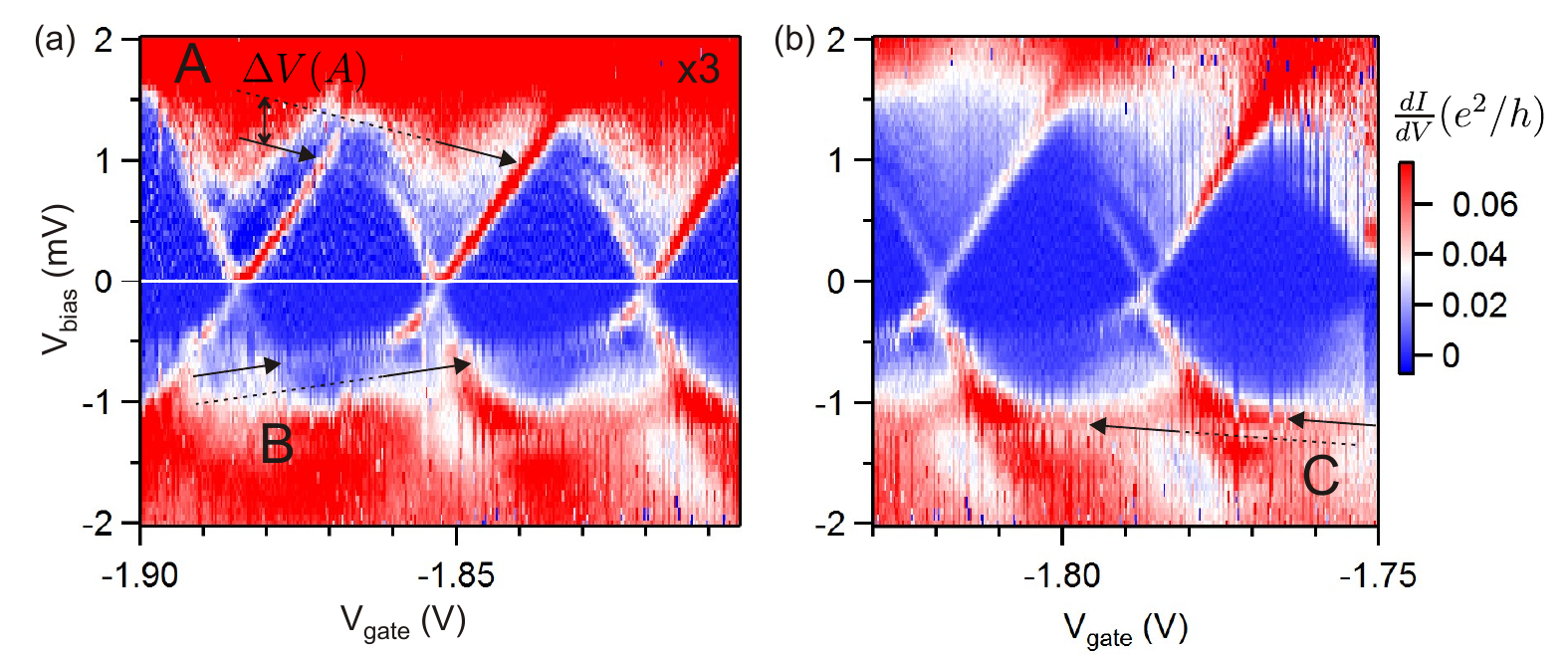}
  \caption{(Color online) Differential conductance plot showing several charge states of the main dot. Secondary resonances of three different side dots $A$, $B$ (in~(a)) and $C$ (in~(b)) are indicated by arrows. The resonances are shifted by $\Delta V(i)$ ($i=A,B,C$) due to a capacitive inter-dot coupling. Note the multiplied differential conductance at positive bias voltage in panel (a).}
  \label{fig:diamonds}
\end{figure*}

This manuscript is structured as follows. After indicating the sample fabrication and measurement techniques in section II, we evaluate in section III the coupling and interaction parameters of four parallel quantum dots within the CNT rope. This section also includes a discussion of the differential gating effect, which allows to tune the system into in-resonance and off-resonance states. Section IV compares master equation calculations with quantum transport measurements in order to show the effect of relative tunnel rates of parallel quantum dots. At last in section V, we briefly outline how to extract the true capacitive coupling between parallel quantum dots in the presence of a strongly asymmetric coupling to one lead.

\section{Sample \& Methods}
We show quantum transport measurements of a CNT rope device with the distance between the gold contacts being patterned as 360\,nm (see Fig.~\ref{fig:afm}). The highly doped Si substrate with SiO$_2$ on top of it acts as backgate. The CNTs are grown on substrate by the chemical vapour deposition method using Fe/Mo as catalyst and methane as the carbon precursor.\cite{Kong1998} The growth temperature is 920\,$^\circ$C, where a predominant growth of single-walled carbon nanotubes and a small fraction of double-walled CNTs is expected \cite{Spudat2009}. The height profile taken from the atomic force micrograph in Fig.~\ref{fig:afm} in the quantum dot region between the two contacts gives a height of $\sim 7$\,nm.  Raman scattering measurements performed on the device reported here, can clearly exclude an individual multiwalled CNT of a large diameter and prove the bundling of several carbon nanotubes with diameters between 0.6\,nm and 1.3\,nm in the quantum dot region.\cite{Goss2011b} The device exhibits a resistance of 290\,k$\Omega$ at room temperature with a linear current-voltage characteristic indicating its metallic character. Low-temperature transport measurements are performed in a dilution refrigerator at a base temperature of $\sim$30\,mK.

\begin{figure}[hb]
  \includegraphics{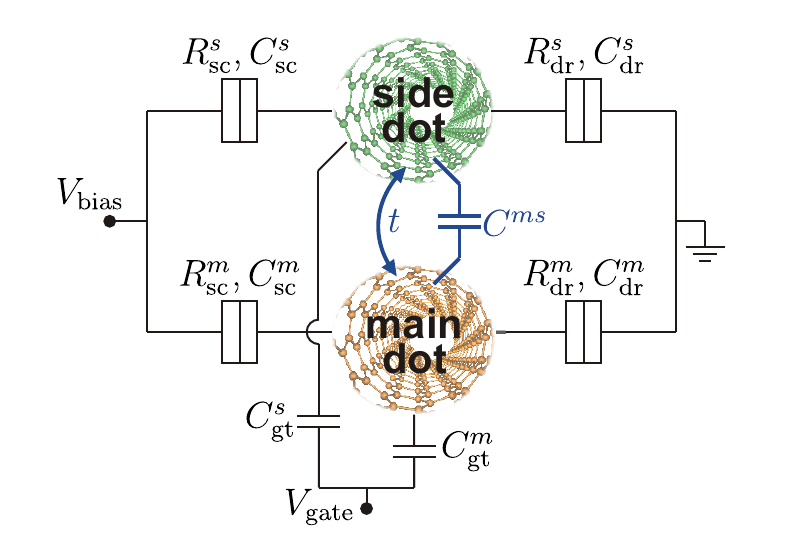}
  \caption{(Color online) Circuit diagram of a system of two parallel quantum dots formed in two CNTs. The tunnel barriers connecting them to the same source (sc) and drain (dr) electrodes are characterized by a resistance $R$ and a capacitance $C$. The quantum dots interact via a tunnel coupling with the hybridization amplitude $t$ and a capacitive coupling with the capacitance $C^{ms}$.}
  \label{fig:circuit}
\end{figure} 

\section{Multiple parallel quantum dots}

Figure~\ref{fig:diamonds} shows the stability diagram of the CNT rope device measured at low temperatures, which exhibits Coulomb diamonds as typical Coulomb blockade signatures and more atypical additional resonances. Previously, we have shown that the observable features can be explained by a formation of interacting parallel quantum dots in different strands of the rope.\cite{Goss2011a} The Coulomb diamonds in Fig.~\ref{fig:diamonds} originate from one quantum dot which we label as main dot and the secondary resonances marked by arrows originate from side dots formed in parallel CNTs within the rope. These secondary resonances are part of a Coulomb diamond pattern with smaller slopes of the diamond edges owing to a weaker backgate coupling of the side dots. 

Within the gate voltage range plotted in Fig.~\ref{fig:diamonds}a and b, indications of three different side dots are observed as we will discuss in the following. Three main signatures are used to discriminate these three quantum dots. First, anticrossings caused by a tunnel coupling between quantum dot states on different dots can be observed at meeting points between resonance lines. This tunnel coupling is characterized by a hybridization amplitude $t$ as sketched in the circuit diagram of the quantum dot system in Fig.~\ref{fig:circuit}. Second, a capacitive coupling characterized by the capacitance $C^{ms}$ between the dots leads to a voltage shift $\Delta V(i)$ in the secondary resonances when proceeding from one Coulomb diamond to the next. This energy shift of a side dot level is caused by a potential change on the side dot due to the subsequent addition of electrons onto the main dot. 

\begin{table}[tb]
\centering
\begin{tabular}{cccccc}
  \hline
   QD $i$ & $\alpha_{\mathrm{gt}}$   & $\alpha_{\mathrm{sc}}$   & $\alpha_{\mathrm{dr}}$ & $|t|$\,(meV) & $\Delta V (i)$\,(meV)\\
  \hline
   main dot   &  0.05               & 0.36                   & 0.59                   & -            & -\\
   side dot~$A$ &  0.019              & 0.765                  & 0.216                  & 0.1          & 0.40\\
   side dot~$B$ &  0.007              & 0.074                  & 0.919                  & 0.075        & 0.20\\
   side dot~$C$ &  0.006              & 0.939                  & 0.055                  & $\neq 0$     &0.15 \\
  \hline
\end{tabular}
  \caption{Coupling and interaction parameters of the quantum dots observed in Fig.~\ref{fig:diamonds}.}
  \label{tbl:alldots}
\end{table}

Third, the bias coupling, that is due to the interaction of the QDs with the leads, can be deduced from the slopes of the diamond edges in the stability diagram. In our experiments, the bias voltage is applied asymmetrically, i.\,e. the drain is kept grounded and the full bias window is applied at the source electrode. Then, the slopes of the diamond edges of each quantum dot are given as $+\frac{\alpha_{\mathrm{gt}}}{1-\alpha_{\mathrm{sc}}}$ and $-\frac{\alpha_{\mathrm{gt}}}{\alpha_{\mathrm{sc}}}$ for the positive and the negative slope, respectively.\cite{Hanson2007} Here, $\alpha_j=C_j/C$ is the dimensionless coupling parameter of the quantum dot to electrode $j$, where $C=\sum C_j$ is the total capacitance as defined for a single quantum dot within the constant interaction picture.\cite{Kouwenhoven2001}

\begin{figure}[htb]
  \includegraphics{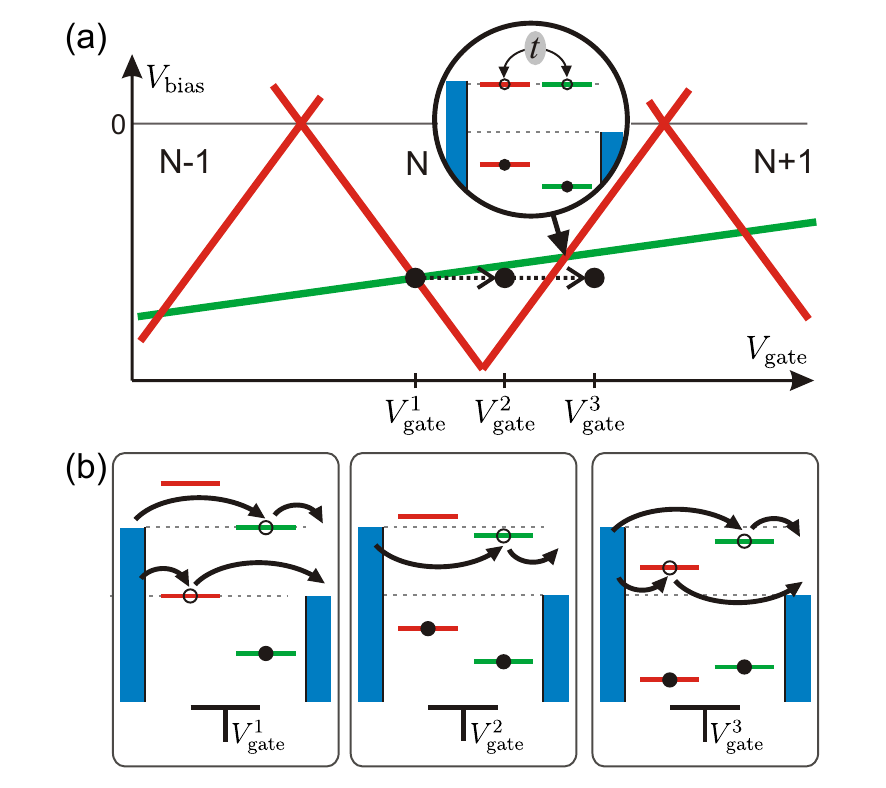}
  \caption{(Color online) (a) Schematic stability diagram of a main dot (red) in parallel to one side dot (green) in the region of negative bias voltage. (b) Level alignment for three gate voltages as indicated in (a) at a constant bias voltage. At $V_{\mathrm{gate}}^1$, tunneling via both quantum dots is possible. Due to the differential gating, it is possible to add one more electron onto the main dot, while the side dot occupancy does not change for $V_{\mathrm{gate}}^2$. At $V_{\mathrm{gate}}^3$, again tunneling via two dots - but different states - is possible.}
  \label{fig:diffgating}
\end{figure}

The coupling parameters $\alpha_j$ deduced from Fig.~\ref{fig:diamonds} are summarized in Tab.~\ref{tbl:alldots}. The parameters of the main dot are directly extracted from the observed diamond pattern. For the side dots, it is not possible to observe complete Coulomb diamonds. Considering each secondary resonance separately, a large bias coupling, i.\,e. a strong coupling to either the source or the drain electrode, has to be considered. In the case of a large source coupling ($\alpha_{\mathrm{sc}} \gg \alpha_{\mathrm{dr}}$), the positive slope becomes very steep, while the negative slope becomes flat. On the other hand, if the coupling to the drain electrode is large, the opposite will be observed: the positive slope is flat, while the negative slope is steep. We interpret each secondary resonance to be the flat slope of a diamond pattern, hence belonging to different side dots, labelled as $A$, $B$ and $C$ in Fig.~\ref{fig:diamonds}. Due to its steepness, the second slope of each side dot diamond is assumed to be obscured by the prominent main diamond pattern. In order to obtain an estimate for the coupling of the side dots to the electrodes, this slope of the side dot diamonds is assumed to be the same as for the main dot. If it was much smaller, it would be observable as a resonance in the diamond pattern. An evaluation with steeper slopes up to completely vertical did not change the results significantly.

The coupling parameters found by this analysis are given in Tab.~\ref{tbl:alldots}. For side dot $A$ and $C$ we find a strong source coupling, while side dot $B$ exhibits a strong drain coupling. The coupling to the gate electrode is similar for side dot $B$ and $C$, while it is slightly stronger for side dot $A$. A screening of particular CNTs by surrounding CNTs within the rope is a possible reason for a different coupling to the backgate. Furthermore, the asymmetric coupling to one of the contacts is a clear indication for different interface properties at the leads, which can originate from a changing assembly of the rope along the QD region.

\begin{figure*}[htb]
  \includegraphics{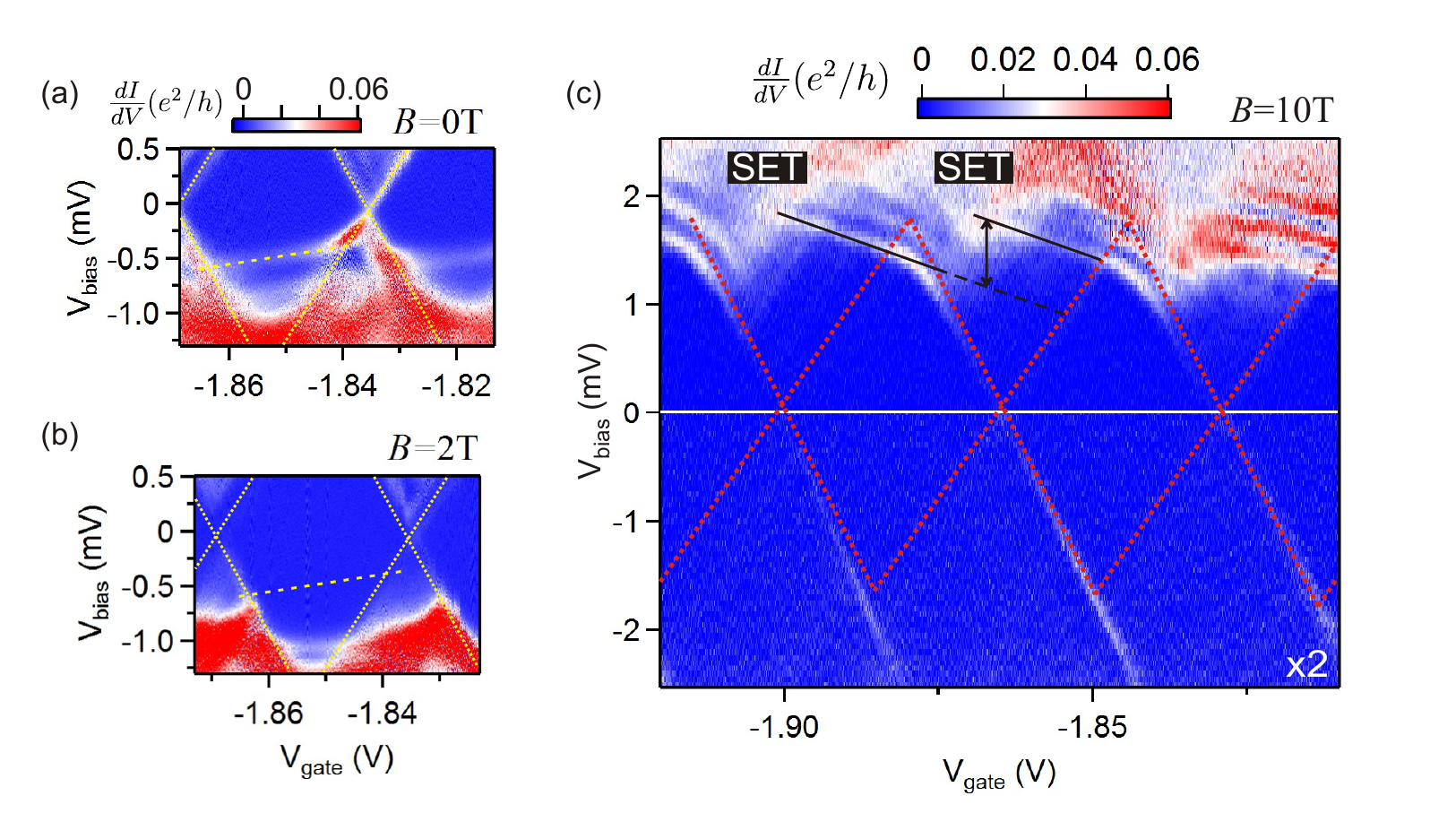}
  \caption{(Color online) Differential conductance plots for magnetic field (a)~$B=0$\,T, (b)~$B=2$\,T and (c)~$B=10$\,T. (b)~The main and secondary resonances as observed at $B=0$\,T are indicated in yellow. (c)~Note the multiplied differential conductance at negative bias voltage. For better visibility, main dot Coulomb diamonds are indicated in red.}
  \label{fig:Bfield}
\end{figure*}

Although all quantum dots in the system are controlled by the same gate electrode, the dot-dependent backgate coupling creates a differential gating effect that is used to tune the system into various states. This is illustrated in Fig.~\ref{fig:diffgating} for one main dot (red) and one side dot (green) where interactions between the QDs are neglected for simplicity. At the first indicated position in the stability diagram in Fig.~\ref{fig:diffgating}a, the bias voltage and the gate voltage $V_{\mathrm{gate}}^1$ are set such that tunneling via both quantum dots is possible simultaneously. The level alignment for this situation is sketched in the left panel of Fig.~\ref{fig:diffgating}b: while the level on the main dot is in resonance with the chemical potential of the drain (negative slope of the Coulomb diamond), the level on the side dot is in resonance with the chemical potential of the source (positive slope of the secondary resonance). Keeping the bias voltage fixed and increasing the gate voltage to $V_{\mathrm{gate}}^2$, lowers the main dot level below the bias window and fills the main dot permanently with one more electron (see central panel of Fig.~\ref{fig:diffgating}b). The main dot is in Coulomb blockade, but tunneling via the side dot level is still possible. This is a particular property of the parallel quantum dot system in contrast to a setup with serial quantum dots. There, the transport via the whole device is fully blocked when only one of the dots is in Coulomb blockade. 

Increasing the gate voltage further to $V_{\mathrm{gate}}^3$, tunes the next higher energy level of the main dot into the bias window allowing again tunneling via both quantum dots in parallel. While the level on the main dot had a higher energy than the side dot level at $V_{\mathrm{gate}}^2$, it exhibits a lower energy at $V_{\mathrm{gate}}^3$, which -- with only one gate electrode -- is solely possible with a differential gating effect. The situation where the electronic states of the two quantum dots are at the same energy, i.\,e. in resonance with each other, is depicted in the inset of Fig.~\ref{fig:diffgating}a. To probe this in-resonance state and a possible hybridization of quantum dot levels, both levels need to be simultaneously in resonance with one of the electrode chemical potentials. From earlier investigations we know that the current is carried by the bonding state of hybridized levels and that the wavefunction overlap is symmetric.\cite{Goss2011a}

Hence, the differential gating effect as it is observed here, can be used to tune the quantum dot system into in-resonance and off-resonance states and provides a control over which strand of the rope carries the current. This enables us to use quantum transport as a spectroscopic tool for probing various properties of the system. 

Now that it is clear how the differential gating effect can be employed to characterize the parallel quantum dot system, we want to use it to distinguish between the side dots by evaluating their interaction parameters in the following. From the discrete shift of a secondary resonance $\Delta V(i)$ between two subsequent main dot Coulomb diamonds, the capacitive inter-dot coupling can be extracted. This shift is found to be twice as large for resonance $A$ in comparison to resonance $B$ (see Tab.~\ref{tbl:alldots}). For side dot $C$, an even smaller capacitive coupling is found. The second interaction is the tunnel coupling to the main dot, as mentioned above. A hybridization between states on two different quantum dots causes an anticrossing between resonance lines. The gap between a bonding and an anti-bonding state observable at the anticrossings corresponds to $\Delta E=2 |t|$, where $t$ is the hybridization amplitude.\cite{Goss2011a} From a high resolution measurement (shown in Ref.~\onlinecite{Goss2011a}), $|t|$ can be estimated to be 0.1\,meV and 0.075\,meV for $A$ and $B$, respectively. For side dot $C$, a magnitude for the hybridization amplitude cannot be accurately determined. However, the clear bending of resonance lines indicate that a hybridization indeed occurs also for these quantum dot states. Hence, the different magnitude of the interactions between the quantum dots gives evidence that Fig.~\ref{fig:diamonds} contains the fingerprints of one main dot connected in parallel to three side dots.


\begin{figure}[htb]
  \includegraphics{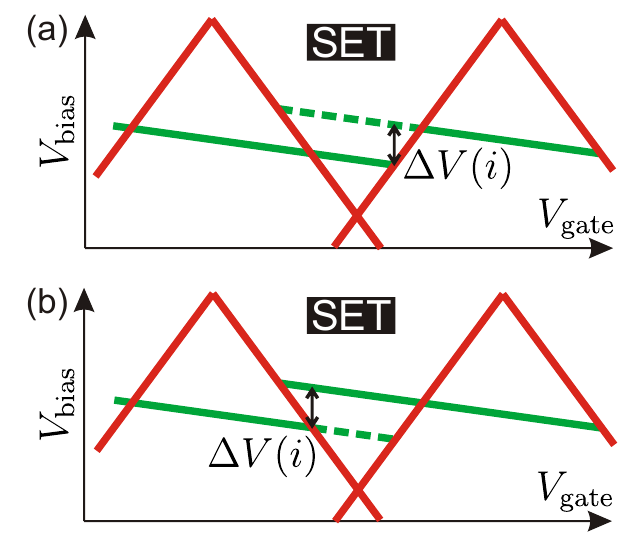}
  \caption{(Color online) Schematic stability diagram of a main dot (red) in parallel to one side dot (green) in the region of positive bias voltage. The voltage shift $\Delta V(i)$ occurs in (a)~at the Coulomb diamond edge and in (b)~in the SET regime of the main dot.}
  \label{fig:compete-sidedot}
\end{figure}

A further evidence for interpreting the secondary resonances as three side dots can be given from a distinct evolution with increasing magnetic field. Figures~\ref{fig:Bfield}a and~b show the stability diagram of the main dot and the secondary resonance $B$ at a magnetic field $B=0$\,T and 2\,T, respectively. While at $B=0$\,T, the secondary resonance is passing the Coulomb blocked region at $V_{\mathrm{bias}} \approx -0.6$\,mV, it appears at a more negative bias voltage for $B=2$\,T. Figure~\ref{fig:Bfield}c presents several charge states of the main dot at $B=10$\,T. At this high magnetic field, no secondary resonances can be observed at negative bias voltages, where the signatures of side dot $B$ and $C$ could be measured at zero magnetic field (see Fig.~\ref{fig:diamonds}). In contrast, enhanced secondary resonances can be observed at positive bias voltage indicating electrons tunneling favourably via side dot~$A$. The contrary evolution of the secondary resonances in a magnetic field confirms once again the presence of several parallel side dots. 

Comparing the secondary resonances $A$ in Fig.~\ref{fig:Bfield}c with those in Fig.~\ref{fig:diamonds}a, two differences can be observed in the single electron tunneling (SET) region of the main dot. First of all, excited states of side dot $A$ are visible, which cause parallel secondary resonances with equal coupling and interaction parameters. A detailed discussion concerning these excited states can be found elsewhere.\cite{Goss2011a} Second, the voltage shift $\Delta V(A)$ of the secondary resonance is found to occur within the SET regime (see the arrow in Fig.~\ref{fig:Bfield}c), and not at the edge of the subsequent Coulomb diamond. This effect is best visible for the secondary resonance with the lowest energy and has its origin in the relative tunneling rates of the main dot and the side dots, which will be discussed in the next section.

In this first section, we showed how a differential gating effect enables the identification and characterization of several parallel quantum dots. A strong bias coupling is found for the side dots in the setup. By tuning to in-resonance and off-resonance states of the system, various interactions can be probed. Furthermore, the differential gating is used to control which one of the strands within the rope carries the current.

\section{Rate-dependent quantum transport features}

\begin{figure}[htb]
  \includegraphics{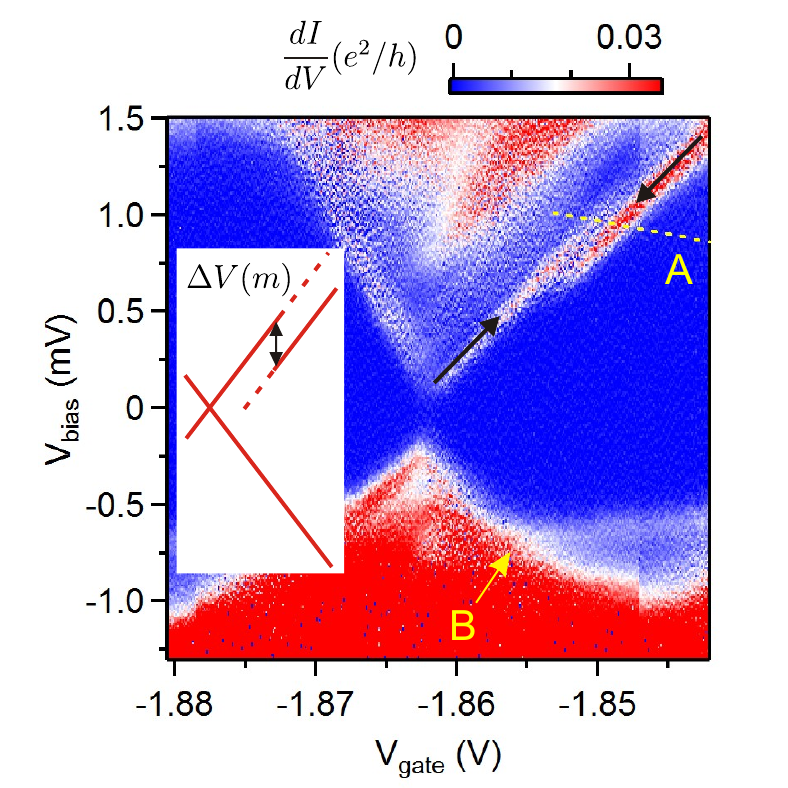}
  \caption{(Color online) Differential conductance plot exhibiting a shifted main resonance (arrows). The secondary resonances of side dot~$A$ and $B$ are indicated in yellow.}
  \label{fig:jump-exp}
\end{figure}

Figure~\ref{fig:compete-sidedot} depicts two possible observations in the SET region of a main dot in parallel to a capacitively coupled side dot (a hybridization of quantum dot states is neglected for simplicity). In the first case (a), the voltage shift $\Delta V(i)$ occurs exactly at the diamond edge of the right diamond, where an additional electron is filled onto the main dot. The second case (b) shows the up-shifted secondary resonance already in the SET regime of the main dot. This observation can be made if an electron resides for a long time on the main dot state within the bias window due to a reduced tunneling rate towards the drain electrode. This is the effect mentioned in section III which can be observed in Fig.~\ref{fig:Bfield}c for the main dot and side dot~$A$. In a magnetic field ($B=10$\,T), the conductance via the main dot is very low (faint Coulomb diamonds) indicating reduced tunneling rates of the main dot. Furthermore, the shift of the secondary resonance is found to occur in the SET regime, and not at the diamond edge. At lower bias voltages, where an unshifted resonance is expected (dotted line), no conductance can be measured, in analogy to Fig.~\ref{fig:compete-sidedot}b.

\begin{figure}[htb]
  \includegraphics{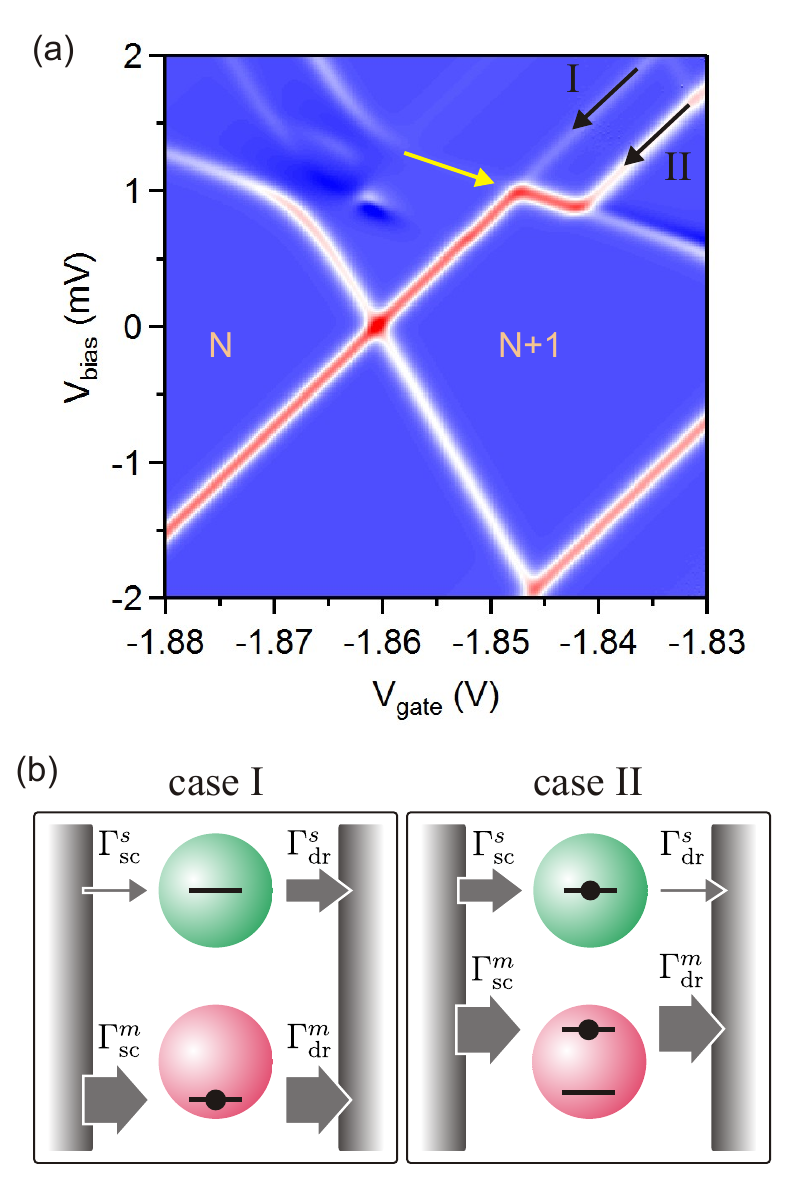}
  \caption{(Color online) (a)~Differential conductance caculated by master equations using the model of two interacting parallel quantum dots. Black arrows indicate the shift of the main resonance appearing at the secondary resonance (yellow arrow). (b)~Schematic drawing of electron tunneling via two parallel CNT quantum dots. $\Gamma_{\mathrm{sc/dr}}^s$ and $\Gamma_{\mathrm{sc/dr}}^m$ are the tunneling rates from source (sc) and drain (dr) to the side dot $s$ (green) and the main dot $m$ (red), respectively. (I)~If $\Gamma_{\mathrm{sc}}^s \ll \Gamma_{\mathrm{dr}}^s$, the side dot is mainly unoccupied and electrons can tunnel via the main dot like in an uncoupled quantum dot. (II)~In the case of $\Gamma_{\mathrm{sc}}^s \gg \Gamma_{\mathrm{dr}}^s$, an additional electron resides on the side dot most of the time and electrons tunneling via the main dot can only tunnel via a state shifted in energy by the capacitive inter-dot coupling.}
  \label{fig:jump-sim}
\end{figure}

A similar effect with the same physical origin can be observed in Fig.~\ref{fig:jump-exp}a, which shows another transport measurement at $B=0$\,T at similar charge states as in Fig.~\ref{fig:diamonds}. The secondary resonance of side dot $B$ is observed similarly to the previous measurement. The position of the secondary resonance from side dot~$A$ is difficult to recognize due to a low conductance in this particular measurement. Instead, a shift of the diamond edge $\Delta V(m)$ can be observed close to the meeting point of the secondary resonance with the main resonance. This is - like the voltage shift in the secondary resonances - a manifestation of the capacitive coupling between the main dot and side dot~$A$. In a system of capacitively coupled quantum dots, the chemical potential of the quantum dots will mutually depend on each other. Hence, resonance lines of any dot -- not only a side dot -- are expected to shift when the number of electrons on the parallel quantum dot is changed. That means, that also the tunneling of an additional electron via a side dot state changes the potential for the electrons on the main dot and a shifted main dot diamond edge is expected at energies higher than the respective secondary resonance, i.\,e. in the single electron tunneling regime of the side dot. This effect causes the observed shift in Fig.~\ref{fig:jump-exp}. As we will see in the following, the relative tunnel rates of quantum dot states are the decisive parameter for a shift of resonances.

We use master equation calculations with an analogous model as in Ref.~\onlinecite{Goss2011a}, describing a parallel double quantum dot (containing the main dot and side dot~$A$) within a constant interaction model \cite{Oreg2000} extended to account for interactions between the quantum dots. Figure~\ref{fig:jump-sim}a shows a calculated stability diagram according to this model. Similar to the measurement, the main dot diamond edge is shifted at the crossing of the secondary resonance. The calculations assume a larger capacitive coupling than observed in the experiment in order to enhance the visibility of the effect.

In the calculated stability diagram, the shifted main dot diamond edge appears only for particular relative tunneling rates from the leads to the quantum dots. The following discussion assumes one state on each quantum dot to be within the bias window and the applied bias voltage to be positive. The left panel (case~I) of Fig.~\ref{fig:jump-sim}b depicts a situation, where the tunneling rates for the state on the side dot are $\Gamma_{\mathrm{sc}}^s < \Gamma_{\mathrm{dr}}^s$, while $\Gamma_{\mathrm{sc}}^m = \Gamma_{\mathrm{dr}}^m > \Gamma_{\mathrm{dr}}^s$ is assumed for the main dot. Then, the side dot state is mainly unoccupied, because electrons will immediately tunnel out of the dot into the drain. Therefore, electrons tunneling via the main dot will not experience an additional potential, because the probability to have an electron on the side dot at the same time is low. As a result, the diamond edge in the stability diagram will appear as a continuous resonance line and exhibit no shift as for uncoupled QDs (see resonance~I in Fig.~\ref{fig:jump-sim}a).

For the calculation shown in Fig.~\ref{fig:jump-sim}a, the tunneling rates for the side dot are assumed to be reversed, i.\,e. $\Gamma_{\mathrm{sc}}^s > \Gamma_{\mathrm{dr}}^s$, as depicted in the right panel (case~II) of Fig.~\ref{fig:jump-sim}b. For the main dot, $\Gamma_{\mathrm{sc}}^m = \Gamma_{\mathrm{dr}}^m > \Gamma_{\mathrm{sc}}^s$ is assumed. This leads to the following situation in the SET regime (i.\,e., above the side dot resonance): the state on the side dot is mainly occupied, and transport predominantly takes place via a state on the main dot, which is shifted in energy due to the inter-dot capacitive coupling. Then, the conductance in the original diamond edge (side dot not occupied) is suppressed, whereas the shifted diamond edge (side dot occupied) appears enhanced, which causes the higher conductance in resonance~II in Fig.~\ref{fig:jump-sim}a. 

This configuration of tunneling rates also suppresses the conductance through the side dot, hence the secondary resonance appears less pronounced. This is consistent with the experimentally found weak conductance of the secondary resonance of side dot~$A$ in Fig.~\ref{fig:jump-exp}.

In conclusion, by reproducing the experimental observations with master equation calculations, we have shown that the energy and the conductance of the resonances in the stability diagram of interacting parallel quantum dots strongly depend on the tunnel rates into particular quantum dot states. Hence, the tunneling rates of one quantum dot dictate the relative conductance for the resonances of the parallel quantum dot and are decisive for the observation of their energy splitting.

\section{Evaluating the true capacitive coupling strength}
As we have shown above, a shift in the resonance lines of the side dot as well as of the main dot is a measure of the capacitive coupling between parallel quantum dots. However, here we find the magnitude of the shift of the main resonance to be $\Delta V(m) \approx 0.2$\,meV, whereas $\Delta V(A) \approx 0.4$\,meV is obtained from the voltage shift of the secondary resonance. This discrepancy can be explained by considering the difference in the bias coupling of side dot~$A$ and the main dot. In fact, the voltage offset due to a capacitive inter-dot coupling and also the magnitude of the anticrossing gap depend on the bias coupling, which we described in section III. A large $\alpha_{\mathrm{sc}}$ will increase the observable shift of resonances, because $\Delta V(i)= U^{ms}/(1-\alpha_{\mathrm{sc}}(i))$, where $\alpha_{\mathrm{sc}}$ is the source coupling of the dot corresponding to the shifting resonance and $U^{ms}$ is the true capacitive coupling strength. The values of the shifts are thus always upper bounds for the actual capacitive inter-dot coupling. 

Considering the source coupling of the main dot and the side dot~$A$ in Tab.~\ref{tbl:alldots}, a shift of 
\begin{equation}
\Delta V(m) = \Delta V(A) \dfrac{1-\alpha_{\mathrm{sc}}(A)}{1-\alpha_{\mathrm{sc}}(m)}=0.15\,\mathrm{meV}
\end{equation}
is expected for the main dot resonance. This value is comparable to the $\Delta V(m)$ deduced from the shift of the main resonance, taking into account the experimental error. In principle, the source coupling of the main dot is more reliable than the ones of the side dots, because no assumptions had to be made for obtaining them. Hence, we evaluate the true capacitive inter-dot coupling between main dot and side dot~$A$ from the shifted main diamond edge and obtain $U^{mA}=0.13$\,meV. 
Although several side dots are interacting with the main dot, only side dot~$A$ is coupled strongly enough to shift the main dot resonances in energy.

\section{Summary \& Conclusions}

In summary, we discussed a multi-component molecular transport device employing quantum transport as a spectroscopic tool. We use a carbon nanotube rope as a molecular model system, which exhibits rich characteristics in its transport spectra. The formation of multiple parallel quantum dots is concluded from secondary resonances. A differential gating effect due to a dot-dependent gate coupling allows for tuning the quantum dot system into various states using only one gate electrode. By tuning to in- and off-resonance states of the system, quantum transport spectroscopy can distinguish several parallel quantum dots by their coupling and interaction properties, as we have shown on three side dots coupled in parallel to one main dot. Furthermore, the differential gating provides a control over the carbon nanotube strand carrying the current.

The distinct coupling of the quantum dots to the leads, their interactions and the impact of relative tunneling rates lead to a variety of features in the transport spectra which we discussed in depth. By reproducing the experimental data with master equation calculations, we have identified the requirements for observing an energy offset due to a possible interaction between parallel quantum dots. Concentrating on the off-resonance quantum transport, we have found that a capacitive inter-dot coupling is not only observable in a shift of secondary resonances, but additionally can manifest itself in shifting the edges of otherwise regular Coulomb diamonds. The exact position and appearance of such a shift depends on the relative tunneling rates through the parallel quantum dots.

Our results provide the basis for understanding quantum transport via parallel quantum dots, which is an important issue in molecular transport where a large variety of hybrid transport devices with new functionalities are expected.

\begin{acknowledgments}
We thank H.\ Schoeller and U. Schollw\"ock for fruitful discussions and S. Trellenkamp for e-beam writing, as well as S.\ Est\'{e}vez Hern\'{a}ndez, H.\ Kertz, H.\ Pfeifer, W.\ Harneit and J.\ Lauer for technical assistance. We acknowledge the DFG (FOR 912) and the European Union Seventh Framework Programme (FP7/2007-2013) under agreement no~270369 (ELFOS) for funding.
\end{acknowledgments}

\end{document}